\documentclass[letterpaper,twocolumn,10pt]{article}
\usepackage{usenix-2020-09}
\usepackage{dsfont}
\usepackage{enumitem}
\usepackage{graphics}
\usepackage{graphicx}
\usepackage{xcolor}

\usepackage{url}
\usepackage{hyperref}
\usepackage{comment}
\usepackage{tabularray}
\usepackage{tabularx}
\usepackage{comment}
\usepackage{lipsum}
\usepackage{mathtools}
\usepackage{cuted}
\usepackage{multirow}
\usepackage{booktabs}
\usepackage{amssymb}
\usepackage{empheq}
\usepackage{caption}
\usepackage{stfloats}

\usepackage[linesnumbered,ruled]{algorithm2e}
\SetKwInput{KwInput}{Input}                
\SetKwInput{KwOutput}{Output}              

\newtheorem{definition}{Definition}

\newlist{contract}{enumerate}{10}
\setlist[contract]{label*=\arabic*.}
\setlistdepth{10} 

\begin{document}
\title{Generalized Attacks on Face Verification Systems}

\author{
{\rm Ehsan Nazari}\\
University of Ottawa\\
enazari@uottawa.ca
\and
{\rm Paula Branco}\\
University of Ottawa\\
pbranco@uottawa.ca
\and
{\rm Guy-Vincent Jourdan}\\
University of Ottawa\\
gjourdan@uottawa.ca
}

\maketitle

\begin{abstract}
Face verification (FV) using deep neural network models has made tremendous progress in recent years, surpassing human accuracy and seeing deployment in various applications such as border control and smartphone unlocking. However, FV systems are vulnerable to Adversarial Attacks, which manipulate input images to deceive these systems in ways usually unnoticeable to humans. This paper provides an in-depth study of attacks on FV systems. We introduce the DodgePersonation Attack that formulates the creation of face images that impersonate a set of given identities while avoiding being identified as any of the identities in a separate, disjoint set. A taxonomy is proposed to provide a unified view of different types of Adversarial Attacks against FV systems, including Dodging Attacks, Impersonation Attacks, and Master Face Attacks. Finally, we propose the ``One Face to Rule Them All'' Attack which implements the DodgePersonation Attack with state-of-the-art performance on a well-known scenario (Master Face Attack) and which can also be used for the new scenarios introduced in this paper. While the state-of-the-art Master Face Attack~\cite{shmelkin2021generating} can produce a set of 9 images to cover 43.82\% of the identities in their test database, with 9 images our attack can cover 57.27\% to 58.5\% of these identifies while giving the attacker the choice of the identity to use to create the impersonation. Moreover, the 9 generated attack images appear identical to a casual observer.

\end{abstract}

\section{Introduction}
Face verification (FV) involves checking whether two face images represent the same person~\cite{Lu_Tang_2015}. In 2014, DeepFace~\cite{6909616}, a model based on convolutional neural networks, achieved human accuracy in FV on Labeled Faces in the Wild (LFW) dataset~\cite{LFWTech}. Since then, neural network models have not only surpassed human accuracy but have also progressed to the point where these technologies are utilized in public safety applications such as border control procedures and also in commercial applications like unlocking smartphones.

Under benign conditions, these models demonstrate excellent accuracy. However, their robustness is questionable when faced with an adversary. These models are prone to inaccurate output predictions when faced with imperceptible or perceptible but natural-looking adversarial input images~\cite{9464957}. Such attacks are classified into (1) Physical Attacks, in which the input to a model is manipulated (e.g. printing a photo, or printing a face in 3D and presenting it to the FV system); and (2) Digital Attacks, in which digital images are manipulated to fool the FV system~\cite{9464957}. In this paper, we focus on a specific type of Digital Attack called an Adversarial Attack that manipulates a face image in such a way that to the human eye the changes are not discernible, but the FV system does not recognize the correct identity. These attacks are imperceptible to humans and can drive the construction of Physical Attacks, thus presenting a high risk. 

Various attacks have been developed to deceive an FV system. For instance, in the Dodging Attack a 
face image is de-identified so the FV system cannot recognize it as the original face, but to the human eye it is still the same person~\cite{8803803,9412236}. Another method is known as the Impersonation Attack, where a face image of an individual A is altered to be recognized as a different desired individual B while, to the human eye, the manipulated image is still recognized as the initial individual A~\cite{Pautov_2019}. A third attack scenario named Master Faces or Master Face Attack, tries to generate images that impersonate a wide number of identities~\cite{nguyen2020generating,nguyen2021master,shmelkin2021generating}. In this paper, we provide a unified view of these attacks; we show that other types of attacks that present a high risk for the FV systems are possible, and we present a novel attack framework capable of easily targeting specific types of these attacks and which outperforms state-of-the-art methods.  


We propose a generalized attack definition, named DodgePersonation Attack, that encompasses the Dodging, Impersonation, and Master Face Attacks while also introducing new attacks.
The DodgePersonation Attack is formulated as follows: we select a set of human face images called the $\mathcal{M}$atchSet, and another disjoint set called the $\mathcal{D}$odgeSet. 
Our objective is to generate one or several input face images to the FV system, that collectively impersonate all identities in the $\mathcal{M}$atchSet, while effectively dodging recognition as any of the identities in the $\mathcal{D}$odgeSet. 
Furthermore, we propose an approach to carry out the DodgePersonation Attack. Moreover, we maintain control over the visual identity of the generated attack images, whereas the previous work did not have such control. Additionally, we aim for these input face images to be indistinguishable to the human eye, appearing as a single image of the same person. For the special case of the Master Face Attack, our results are significantly better than previous work~\cite{shmelkin2021generating}. In addition, we solve new attack scenarios emerging from the DodgePersonation Attack using the same framework. 

Our key contributions are as follows:

\begin{contract}
    \item We provide a holistic view of attacks against FV systems by defining the DodgePersonation Attack, a generic attack that integrates various attack types. 
    \item We propose a taxonomy of different Adversarial Attacks against FV systems based on their target.
     \item We introduce a novel algorithm named ``One Face to Rule Them All" to deploy the DodgePersonation Attack on FV systems that exhibits state-of-the-art results while allowing complete control over the identity of the generated attack images. 
\end{contract}

In the rest of the paper, we review the relevant work in this domain in Section~\ref{Background}, define the DodgePersonation Attack in Section~\ref{Problem}, and present our new ``One Face to Rule Them All" method in Section~\ref{Method}. We discuss our experimental settings in Section~\ref{Evaluation} and present and discuss our results in Section~\ref{Results}. Finally, Section~\ref{sec:conc} concludes the paper. 

\section{Related Work and Problem Motivation} \label{Background}

\subsection{Face Verification Systems}
In general, automatic face recognition can be separated into two main categories: face identification (also known as closed-set face recognition) and face verification. The former involves classifying faces into a pre-determined set of identities, while the latter determines if two unseen identities during the training phase, represent the same person~\cite{Lu_Tang_2015,Dong_2019_CVPR,amada2021universal}. Therefore, FV is a zero-shot learning task, as the identities of the individuals are not known during the training phase~\cite{amada2021universal}. This makes FV a more difficult and complex task when compared to face identification. Attacking an FV system is the focus of this study.

\begin{figure*}[htpb]
    \centering
    \includegraphics[width=0.99\textwidth]{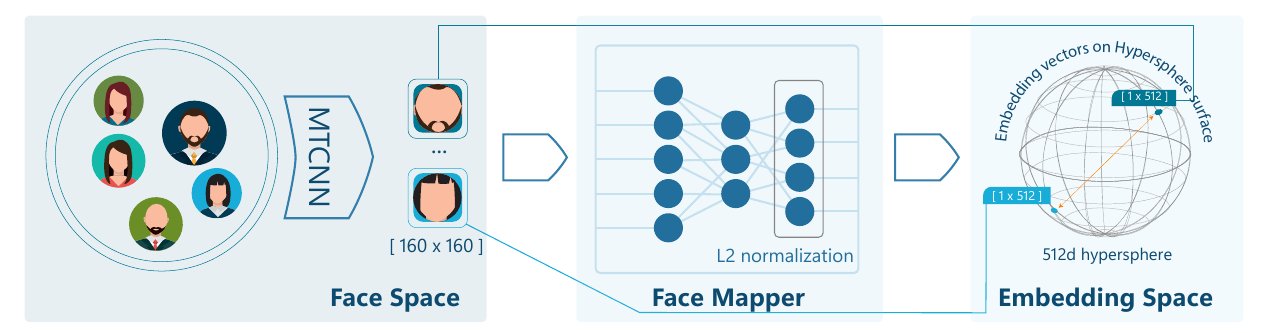}
    \caption{Three-step FV process. First step: automatic detection of faces in a given image using the MTCNN method. Second step: map the faces from the image space to the embedding space using a neural network FM. 
    Third step: face matching by calculating the distance between two face's embeddings and determining if they represent the same identity based on a threshold.}
    \label{fig:FV_System}
\end{figure*}

As shown in Figure~\ref{fig:FV_System}, current state-of-the-art solutions offer a three-step process for FV: i) face detection; ii) face mapping from image space to embedding space; and iii) face matching. For the first step, a commonly used method to detect faces automatically in a given image is the Multi-task Cascaded Convolutional Neural Networks (MTCNN)~\cite{7553523}. For the second step, a face mapper (FM) or face descriptor is used to map a face from the image space into a point in the embedding space. We refer to this point as the embedding corresponding to a face that is processed by an FM. The embedding space is a high-dimensional space where the feature vectors of the faces are projected. Usually, the output feature vector is L2-normalized. Current state-of-the-art FV systems are based on neural network FMs. Finally, in the third step corresponding to the face matching, the distance (e.g., Euclidean distance) between the embeddings of two faces is calculated. Two faces represent the same identity if the distance is below a given threshold. Otherwise, the faces are considered to be of two different persons, i.e., they belong to two different identities.
            
One of the key factors in the formation of an embedding space is the loss function used for training the FM. DeepFace~\cite{6909616}, one of the earliest solutions, used a vanilla softmax loss function. In addition to the three-step FV process mentioned above, it includes a step known as face frontalization after face detection. The method achieved 97.35\% accuracy on the LFW dataset, achieving similar accuracy to human level (97.53\%). Another early method, DeepID2~\cite{NIPS2014_e5e63da7}, and its extension DeepID2+~\cite{sun2015deeply}, improved upon DeepFace by employing a combination of softmax loss and a distance metric to minimize the distance between the same identity pairs and maximize the distance between different identity pairs. These methods achieved an accuracy of 99.15\% and 99.47\% respectively on the LFW dataset. A breakthrough in the field was the introduction of FaceNet~\cite{7298682}, which used a triplet loss function for direct learning of an embedding space. This method further improved the performance, with accuracy reaching 99.63\% on the LFW dataset. In the years following FaceNet's release, several other methods have been proposed to improve upon it. One such method~\cite{NIPS2016_6b180037} generalizes the triplet loss by allowing joint comparison among more than one negative example. Other methods, such as SphereFace~\cite{liu2017sphereface}, CosFace~\cite{wang2018cosface,wang2018additive,8014985}, and ArcFace~\cite{8953658}, have used a combination of softmax loss and a marginal penalty loss to improve performance. 

\subsection{Attacks on Face Verification Systems}
One of the key challenges in FV is the robustness of the methods against various types of attacks. These attacks can take the form of Physical Attacks, such as using masks or makeup to alter the appearance of a face, or Digital Attacks, such as manipulating images or videos~\cite{9464957}. An Adversarial Attack is a type of Digital Attack that involves adding small perturbations to an image in order to fool a model -- in our case, an FV system. Adversarial Attacks can be categorised based on their specificity to non-targeted and targeted~\cite{9464957}. A non-targeted attack, also known as Dodging Attack, is designed to cause an FV system to fail to recognize the correct identity. Research in this area has also focused on methods to generate such attacks, such as~\cite{dong2019efficient,goodman2020advbox,sharif2019general}. A targeted attack, also known as Impersonation Attack, is designed to cause an FV to misidentify a specific individual. The research community has also focused on methods to generate these attacks, such as~\cite{dong2019efficient,zhong2020towards,8803803,goodman2020advbox,sharif2019general}. Additionally, a recently emerging type of Adversarial Attack known as Master Face Attack is designed to cause an FV system to accept the input as a match to all identities, essentially serving as a face master-key. In~\cite{amada2021universal} such images are crafted by minimizing the distance of a perturbed image embedding from a mini-batch of dataset embedding. The authors take an arbitrary face image as input and then craft the attacked image based on it. Finally, three works have been published in recent years that generate Master Faces by searching through the latent space of a pre-trained Generative Adversarial Network (GAN)~\cite{nguyen2020generating,nguyen2021master,shmelkin2021generating}. Unlike in~\cite{amada2021universal}, these works do not have control over the image identity of the synthesized master face, i.e., the attack is not built based on an identity of their choice. Despite being considered Digital Attacks, these are not Adversarial Attacks since they do not alter a given face image by adding perturbations, but create a face that does not exist in reality with the aid of GANs.

The field of FV systems and attacks has some gaps and unresolved issues in the existing related work. Research has mainly focused on individual attack types, neglecting potential connections among them. Our research aims to address this by proposing a holistic view of different attacks, consolidated into a singular problem. This new formulation has led to new attack scenarios not previously considered by the research community. Additionally, we propose a new flexible method capable of conducting various attack scenarios, with success rates that surpass the state-of-the-art.

\subsection{Problem Motivation}


As highlighted earlier, the research conducted in the field of FV systems has primarily revolved around exploring novel attack solutions. The existing body of work has primarily centered on the isolated investigation of individual attack types, thereby overlooking potential interconnections. This limited focus has hindered the development of a comprehensive understanding of the broader attack landscape.

\textbf{DodgePersonation Attack.} To bridge this gap, we propose an attack formulation that \textbf{integrates distinct attack types, offering a unified perspective of FV attacking scenarios}. This formulation consolidates a diverse set of attacks that have been addressed in isolation into a singular definition, with each instance representing a distinct manifestation of an attack. Our approach uses two distinct sets of face images, namely $\mathcal{M}$atchSet and $\mathcal{D}$odgeSet. The primary aim is to deceive the FV system by presenting one or multiple inputs that are collectively recognized as identity matches for individuals in $\mathcal{M}$atchSet, while concurrently evading identification as any member of $\mathcal{D}$odgeSet.

The fact that we might want to match a set of identities seems more intuitive than the condition of having to evade another set of identities. We illustrate why this case is indeed relevant through a recent case in the news. A company allegedly used facial recognition to prevent a lawyer from entering their venues because that lawyer's firm was representing clients involved in a litigation against that company\footnote{Source: \href{https://arstechnica.com/tech-policy/2023/01/msg-probed-over-use-of-facial-recognition-to-eject-lawyers-from-show-venues/}{MSG probed over the use of facial recognition to eject lawyers from show venues}}. This serves as an example of an attack scenario where our proposed system would bypass such restrictive measures. In this scenario, the targeted lawyer, who is denied access, could input their own face into our system and add it to the $\mathcal{D}$odgeSet. The system would generate an output image that resembles the attacker's face but remains unrecognized by the facial recognition system. Additionally, the targeted lawyer has the option to include both their own face and the faces of their coworkers in the $\mathcal{D}$odgeSet. This serves the purpose of not only concealing their own identity but also avoiding being identified as any other individuals who they suspect may also be denied access.


We introduce the concept of \textbf{Source Face}, which represents the user-input face image used to initiate the attack. Carrying out an attack involves generating one or multiple variations of the Source Face, referred to as \textbf{Attack Faces}, for deceiving the FV system and meeting the attack constraints. These image variations are intentionally crafted to be nearly imperceptible to the human eye.
The minimum number of Attack Faces required from a given Source Face depends on the specific attack scenario (which we will discuss in Section~\ref{scenarios}) and the quality and performance of the FM used. For instance, in the case of conducting a Master Face Attack, the more advanced the FM, the greater the number of Attack Faces required to cover most of the provided identities. In practice, we have found that even with a relatively small number of Attack Faces, a significant majority of the identities can be covered. We elaborate on this aspect in greater detail in the subsequent sections.


\subsection{Threat Model}
\subsubsection{Attacker's Objective}
The attacker's objective is to generate one or multiple face images that match as many identities as possible from the $\mathcal{M}$atchSet, while successfully evading detection as any member of the $\mathcal{D}$odgeSet. The attacker might have secondary objectives, such as generating visually identical face images or generating images visually similar to pre-selected ones.

\subsubsection{Attacker's Capabilities and Constraints}
In this study, we consider a white-box attack~\cite{9464957} scenario where the attacker has unrestricted access to the target model. However, the attacker is limited by the requirement to create one or several human-like face images (Attack Faces) that can successfully pass the face detection stage of the FV process. Consequently, the attacker must generate images that meet their goal while still looking like human faces. Furthermore, the attacker can be limited to crafting the face images to closely resemble a specific user-input identity (i.e., the Source Face).

\subsubsection{Possible Attack Scenarios}
Varying populations of $\mathcal{M}$atchSet and $\mathcal{D}$odgeSet, result in distinct attack scenarios. As an illustration, consider the scenario where a wanted criminal, acting as an attacker, intends to fraudulently apply for a license using an online form. Their objective is to impersonate another individual and successfully evade recognition. To achieve this, the attacker needs to generate an image or set of images that, in the eyes of the FV system, resemble the victim's face and differ from their own, while to the human eye, these images appear to be of the attacker. This can be represented as a $\mathcal{M}$atchSet containing the victim's face image and an $\mathcal{D}$odgeSet containing the attacker's face image, with the attacker's face image serving as the Source Face. More scenarios are detailed in Section~\ref{scenarios}.

\subsubsection{Targeted Attack Model}
We utilize FaceNet, a neural network-based FV system, as the target for our attack.


\section{The \textit{DodgePersonation Attack} }\label{Problem}

In this section, we propose a taxonomy for FV system attack scenarios. We first define some terms and notations. Then, we introduce the concepts of impersonation and dodging for face images and define them mathematically. Finally, we introduce two sets of face images: $\mathcal{M}$atchSet and $\mathcal{D}$odgeSet, which form the foundational pillars of the DodgePersonation Attack. These sets are instrumental in defining the problem and serve as a basis for categorizing various attacks within our proposed taxonomy.

\begin{figure*}[htpb]
    \centering
    \includegraphics[width=0.99\textwidth]{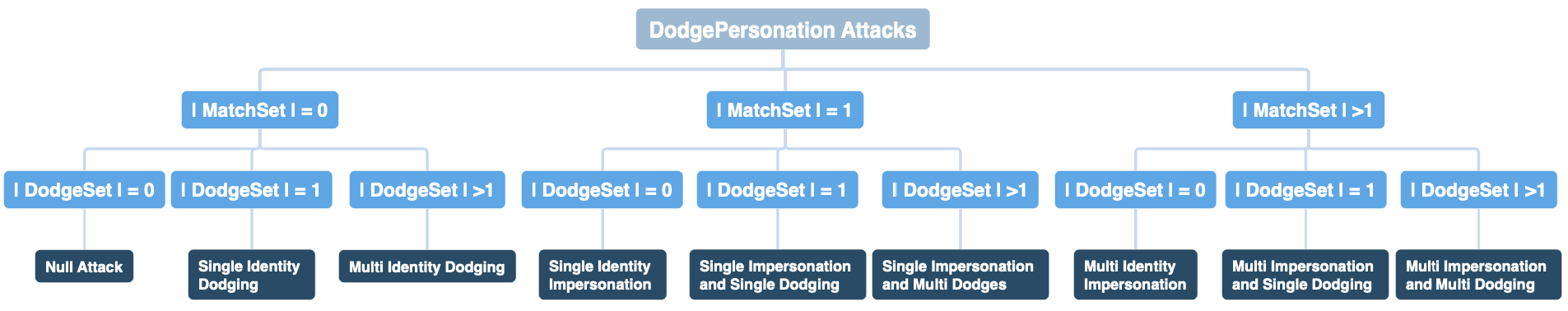}
    \caption{DodgePersonation Attack Taxonomy proposed for categorizing face verification system attack scenarios.}
    \label{fig:taxonomy}
\end{figure*}

\subsection{DodgePersonation Attack Definition and Taxonomy}\label{sub:tax}

Let us assume that $\mathcal{F}$ is the domain of all face images and $\mathcal{I}$ is the set of all identities.
Let $Ident(): \mathcal{F} \rightarrow \mathcal{I}$ represent a function that maps a face image into its identity. If we consider, for instance, three face images, $\{i_1, i_2, i_3\} \subset \mathcal{F}$ of the same identity $I_1 \in \mathcal{I}$, then $Ident(i_1)=Ident(i_2)=Ident(i_3)=I_1$.

Let us assume that a face image is represented by 3 squared matrices $M =m \times m$, each representing a color channel. Consider a given FV system, that is decomposed into two elements: (i) a face mapper function that we represent by $FM()$; and (ii) a distance function that we represent by $Dist()$. The system uses $FM(): M^3 \rightarrow \mathbb{R}^p$ to project a given face image into the embedding space. When an image $x$ (or set of images $S$) is mapped from the face image space into the embedding, it will be represented as $\Bar{x}$ (or $\bar{S}$). This notation will be used for all objects in the embedding space. $Dist(): \mathbb{R}^p \rightarrow \mathbb{R}$ is used to calculate the distance between two points in the embedding space as a hard similarity measure. The matching or mismatching of the identities of two face images in the embedding space is decided using a threshold $th \in \mathbb{R}$. As described in Section~\ref{Background}, $FM()$ uses a neural network architecture to extract features from the images.

\begin{definition}[Impersonation and Dodging]\label{def:1}
Let $A,B \in \mathcal{F}$. We say Face Image $A$ \textbf{impersonates} Face Image $B$ if:
\begin{equation}
      Dist(FM(A),FM(B)) \leq th 
\end{equation}

If the condition is not met, we say Face Image $A$ \textbf{dodges} Face Image $B$.
\end{definition}

\begin{definition}[$\mathcal{M}$atchSet and $\mathcal{D}$odgeSet Face Image Sets]
Let $\mathcal{M}\text{atchSet} = \{m_1, m_2, \cdots, m_{k-1}, m_k \} \subset \mathcal{F}$ represent a set of images, and let $\mathcal{D}\text{odgeSet} = \{d_1, d_2, \cdots, d_{l-1}, d_l \} \subset \mathcal{F}$ represent another set of images. In the rest of the paper, we will assume that sets of face images for a given problem are referred to as meeting certain constraints. Specifically, these sets must fulfill the following requirements:
\begin{itemize}
    \item $k \geq 0$
    \item $l \geq 0$
    \item $\mathcal{M}\text{atchSet} \cap \mathcal{D}\text{odgeSet} = \emptyset $
\end{itemize}

\end{definition}



Having $\mathcal{M}$atchSet and $\mathcal{D}$odgeSet defined, we can now specify the DodgePersonation Attack and categorize the different attacks into a taxonomy.




\begin{definition}[DodgePersonation Attack]\label{dodgepersonation}
Consider two sets $\mathcal{M}\text{atchSet} 
$ and $\mathcal{D}\text{odgeSet} 
$. The DodgePersonation Attack is defined as the following multi-objective optimization problem. The objective is to produce a collection of images called "Attack Faces", represented as $\mathbf{X} = \{x_1, x_2, \ldots, x_n\}$. These images must conform to the requirement of being identified as faces by the MTCNN face detector allowing them to proceed to the subsequent FV steps successfully.  
For $x_i\in X$, let $M_{x_i}$ and $D_{x_i}$ be defined as:


\begin{equation}
\begin{cases}
M_{x_i} := \{ m \in \mathcal{M}\text{atchSet}: Dist(FM(x_i),FM(m)) \leq th \} \\
D_{x_i} : = \{ d \in \mathcal{D}\text{odgeSet}: Dist(FM(x_i),FM(d)) \leq th \}
\end{cases}
\end{equation}
Then, our multi-objective optimization problem is defined as:
\begin{equation}
\begin{cases}
\max \lvert \bigcup\limits_{\forall x_i \in X} M_{x_i} \rvert\\
\min \lvert \bigcup\limits_{\forall x_i \in X} D_{x_i} \rvert
\end{cases}
\end{equation}

\end{definition}

\subsection{DodgePersonation Attack taxonomy} \label{scenarios}

Based on the cardinality of $\mathcal{M}$atchSet and $\mathcal{D}$odgeSet, the DodgePersonation Attack can be divided into multiple types of attacks. These types are displayed in our proposed taxonomy in Figure~\ref{fig:taxonomy}. The three top branches on the first layer characterize the number of identities to be impersonated (i.e., they correspond to different $\mathcal{M}$atchSet sizes). At this level, we can consider the three following scenarios: (1) no impersonation; (2) a single identity impersonation; and (3) multiple identities impersonation. In the second layer, we consider the number of identities to dodge from, which is related to the $\mathcal{D}$odgeSet size. This layer consists of three branches: (1) no identity to dodge from; (2) a single identity to dodge from; and (3) multiple identities to dodge from. Depending on the desired number of identities to impersonate and dodge, nine possible scenarios emerge from this taxonomy. 

The research community commonly refers to scenarios where the $\mathcal{M}$atchSet size is zero and $\mathcal{D}$odgeSet size is one as None-Targeted Attack, Face De-Identification, or Dodging Attack, and where the $\mathcal{M}$atchSet size is one and $\mathcal{D}$odgeSet size is zero as Targeted Attack~\cite{9464957}. Large $\mathcal{M}$atchSet sizes, i.e., containing a large number of identities, where the $\mathcal{D}$odgeSet size is zero, are referred to as Master Faces or Master Face Attack~\cite{nguyen2020generating}. Our proposed taxonomy also includes novel scenarios with no established names in the literature because they have not yet been considered.

Taking a closer look at FV systems from the perspective of an attacker, we can explore the conditions under which the systems can be deceived. The objective of the attacker is to create a set of images that \emph{look like them} to the human eye (i.e., the Source Face is a face image of the attacker) but are identified as a different identity or identities by the FV system. Below, we present a list of real-world scenarios that correspond to specific examples of the cases in our proposed taxonomy, shown in Figure~\ref{fig:taxonomy}.

\textbf{Null Attack.} 
The first scenario, which we call the Null Attack, occurs when there is no intention to impersonate or dodge any identity. This is a trivial case where any valid input is a solution.

\textbf{Single Identity Dodging.} 

In a specific attack scenario, an individual aims to shield their identity from being identified by an online FV system used on social media, thereby seeking to evade the system's facial identity-check mechanism. To do so, the attacker needs to create an image that does not match their face. This scenario can be accomplished by creating an empty $\mathcal{M}$atchSet and a $\mathcal{D}$odgeSet that contains the attacker's own face image. This scenario is named None-Targeted Attack, Face De-Identification, or Dodging Attack by the research community~\cite{9464957}.

\textbf{Multi Identity Dodging.} In this scenario, the attacker, who is a wanted criminal, aims to conceal their identity and ensure that the altered image does not resemble any other potentially sensitive identities, such as other criminals. To minimize the likelihood of being recognized as either themselves or another criminal, the attacker must generate an image that deviates from multiple identities, including their own. This can be achieved by defining a $\mathcal{M}$atchSet with no members and a $\mathcal{D}$odgeSet consisting of the images of those identities.

\textbf{Single Identity Impersonation.} 

In another attack scenario, an attacker may wish to gain access to someone else's smartphone by impersonating the victim's identity~\footnote{In this given scenario, we consider having direct access to the smartphone's FV system. Moving forward, a logical progression for this research would involve exploring Physical Attacks, which represent a more realistic setting for this particular scenario.}. To achieve this, the attacker would need to create a set of images that match the victim's face, which can be formulated as a $\mathcal{M}$atchSet with one member being the face image of the victim with an empty $\mathcal{D}$odgeSet. The research community has given the name Targeted Attack to this scenario, as documented in~\cite{9464957}.

\textbf{Multi Identity Impersonation.} 
This situation can arise when the attacker aims to deceive the online FV system of a portal in order to obtain unauthorized access. In this scenario, the attacker possesses incomplete knowledge regarding the authorized employees, lacking awareness of which employees have access privileges and which ones do not. To increase the chances of success, the attacker needs to impersonate all possible individuals who might have proper access. This scenario can be accomplished by creating a $\mathcal{M}$atchSet that contains several members, which are the face images of the employees. In a similar scenario, the attacker may want to gain access to any arbitrary smartphone by fooling its FV system. To achieve this, the attacker needs to create a set of images that look like a large number of identities. This scenario can be formulated as a $\mathcal{M}$atchSet containing face images of \emph{a large number of identities}, while keeping $\mathcal{D}$odgeSet empty. This scenario is an extreme case of the previous one and is known as a Master Face Attack~\cite{nguyen2020generating}.


\textbf{Single Impersonation and Single Dodging.} 
Another scenario arises when we need to satisfy the requirements of both \textbf{Single Identity Dodging} and \textbf{Single Identity Impersonation} \textbf{at the same time}. In a hypothetical scenario, we can envision a situation where a wanted criminal, acting as an attacker, intends to utilize their own image on an online system equipped with an FV mechanism to gain entry. In this case, the attacker seeks to access the system by impersonating an authorized individual. However, the attacker also wants to make sure that their own identity is not recognized. To achieve this, the attacker needs to create a set of images that match the victim's face and do not match their own face. This can be formulated as a $\mathcal{M}$atchSet with one member being the face image of the victim and an $\mathcal{D}$odgeSet with the attacker's own face image. In the same vein, when the attacker intends to hide their identity and prevent the modified image from resembling any other critical identities, the scenario of \textbf{Single Impersonation and Multi Dodging} arises. Additionally, in situations where the attacker has limited information about the individuals they want to impersonate and is unsure of who has access or not, and also needs to avoid being recognized as themselves, the scenario of \textbf{Multi-Impersonation and Single Dodging} arises.


\textbf{Multi Impersonation and Multi Dodging.} Lastly, if an organization allows access to certain individuals, but alerts the police if certain wanted individuals are recognized, the attacker might want to impersonate the authorized individuals and avoid being recognized as one of the wanted individuals. To increase their chance of gaining access, they can use a $\mathcal{M}$atchSet that contains face images of several authorized individuals who might have access, while using a $\mathcal{D}$odgeSet that contains face images of the wanted individuals. This way, the attacker can minimize the risk of getting caught while maximizing the chances of entering the facility.

\section{One Face to Rule Them All Method} \label{Method}


This section presents the \textbf{One Face to Rule Them All Algorithm} that is used to perform a DodgePersonation Attack. Our algorithm consists of two phases: Phase 1 - the embedding space search; and Phase 2 - the Attack Face generation. In Phase 1, we map the $\mathcal{M}$atchSet and $\mathcal{D}$odgeSet into the embedding space $\mathcal{E}$ using the $FM$ and employ a genetic algorithm (GA) with a special-purpose fitness function to search for a point in the embedding space that satisfies the attack's constraints. During Phase 2, we  modify the user's input face image (the Source Face) to ensure that when it is mapped to the embedding space using $FM$, it is positioned in close proximity to the point discovered in Phase 1. This way, the manipulated user input image satisfies the constraints of the $\mathcal{M}$atchSet and $\mathcal{D}$odgeSet. 

We must highlight that in our algorithm we introduce an additional constraint to the DodgePersonation Attack in Definition~\ref{dodgepersonation}. We require that the set of Attack Faces obtained, i.e., the manipulated images generated by our algorithm, resemble a single~\footnote{While our Algorithm allows for the use of different Source Faces, we consider a special case where all Attack Faces are crafted from a single user-input Source Face. We intentionally confine ourselves to a Single Source Face for all Attack Faces to demonstrate the method's capabilities in crafting nearly identical face images to the human eye. 
As a result, this constraint increases the complexity of the attack generation when compared to the scenario without this restriction.} predetermined identity recognizable to humans (referred to as the Source Face).

\subsection{Phase 1 - Embedding Space Search}\label{sec:4.2}

\begin{figure*}[htpb]
    \centering
    \includegraphics[width=0.99\textwidth]{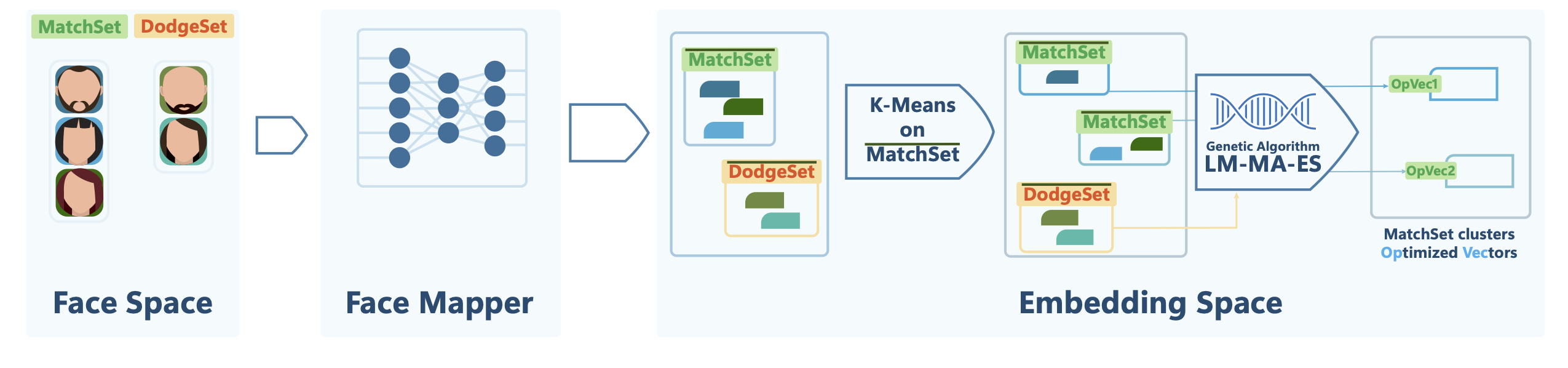}
    \caption{Overview of One Face to Rule Them All Phase 1: Embedding Space Search. The $\mathcal{M}$atchSet and $\mathcal{D}$odgeSet are mapped from the face space into the embedding space. Then, C clusters are built using $\overline{\mathcal{M}\text{atchSet}}$. For each cluster, one point is sought using our genetic algorithm such that it is close to the cluster members and distant from the $\overline{\mathcal{D}\text{odgeSet}}$ members.}
    \label{fig:phase1&2}
\end{figure*}

\paragraph{\textbf{Overview}}
The goal of Phase 1 is to obtain one or more points in the embedding space that can correspond to the Attack Face(s) we want to find. To achieve this we map the $\mathcal{M}$atchSet and $\mathcal{D}$odgeSet using the FV system from the face space into the embedding space where we carry out a search to determine the desired points. Recall that we use a bar on top of the elements that are represented in the embedding space, thus, after this step, we will have $\overline{\mathcal{M}\text{atchSet}}$ and $\overline{\mathcal{D}\text{odgeSet}}$. We propose two key steps for determining the desired points: (i) constructing clusters over $\overline{\mathcal{M}\text{atchSet}}$; and (ii) searching the embedding space around each cluster using a GA with a specially developed fitness function. Figure~\ref{fig:phase1&2} provides an overview of Phase 1.

\paragraph{\textbf{From Face Space to Embedding Space.}}
The faces in $\mathcal{M}$atchSet and $\mathcal{D}$odgeSet are passed through the MTCNN algorithm, normalized, and fed into the $FM$ function, as shown in Figure~\ref{fig:FV_System}. 

\paragraph{\textbf{Cluster Generation.}}                
The final objective is to locate a set of points, $\overline{x_{i}}$, in the embedding space $\mathcal{E}$ that satisfy two conditions: (i) collectively are near all members of the $\overline{\mathcal{M}\text{atchSet}}$
; and (ii) $\overline{x_{i}}$ are far from all members of $\overline{\mathcal{D}\text{odgeSet}}$. 
However, a single point may not meet the requirements since the points in $\overline{\mathcal{M}\text{atchSet}}$ might be dispersed throughout the embedding space. Therefore, we create C clusters of the $\overline{\mathcal{M}\text{atchSet}}$ cases ($\overline{\mathcal{M}\text{atchSet}_1},$ $\cdots,$ $\overline{\mathcal{M}\text{atchSet}_C}$), by applying 
the K-Means algorithm~\cite{1056489}. The number of clusters is a hyper-parameter of the system.

\paragraph{\textbf{Genetic Algorithm and Proposed Fitness Function.}}
For each cluster $\overline{\mathcal{M}\text{atchSet}_i}$, we want to solve the following multi-objective optimization problem: 


\begin{empheq}[left={\empheqlbrace }]{align}
    & max \lvert \{ \overline{m} \in \overline{\mathcal{M}\text{atchSet}}_i: Dist(\overline{x}_{i} ,\overline{m}) \leq th  \} \rvert  && \label{eq:multiobj2a}\\
    & min \lvert \{ \overline{d} \in \overline{\mathcal{D}\text{odgeSet}}: Dist(\overline{x}_{i}, \overline{d}) \leq th \} \rvert && \label{eq:multiobj2b}
\end{empheq}


We employ the LM-MA-ES~\cite{8410043} GA to identify a point in the high-dimensional embedding space satisfying our requirements. We selected this GA because it was shown to be effective for searching high-dimensional spaces~\cite{shmelkin2021generating}. 

We evaluate the effectiveness of a point in meeting the requirements by minimizing our proposed fitness function, the DodgePersonation Fitness function (cf. Definition~\ref{def:ga}), consisting of a positive and a negative component that corresponds to the objectives in Equations~\ref{eq:multiobj2a} and \ref{eq:multiobj2b}. 
The DodgePersonation Fitness function is a weighted sum of these two components that are based on the $DPloss$ (cf. Definition~\ref{def:DodgeCount_DistSum}) that is applied to either the $\overline{\mathcal{M}\text{atchSet}_i}$ or $\overline{\mathcal{D}\text{odgeSet}}$. 

\begin{algorithm}[htpb]
\footnotesize
\caption{One Face to Rule Them All - Phase 1: Embedding space Search}\label{alg:phase2}
  \KwInput{$\overline{\mathcal{M}\text{atchSet}}$: set of all face embeddings to impersonate;\\
  \hspace{0.8cm}$\overline{\mathcal{D}\text{odgeSet}}$: set of all face embeddings to dodge;\\
  \hspace{0.8cm}$\alpha$, $\beta$: weights for positive and negative $DPloss$, respectively;\\
  \hspace{0.8cm}$\gamma$: weight for the DodgePersonation Fitness function\\
  \hspace{0.8cm}$th1$, $th2$: DodgePersonation Fitness thresholds\\
  \hspace{0.8cm}$gen$: number of GA generations\\
  \hspace{0.8cm}$C$: number of clusters 
  }
  \KwOutput{$best\_emb$: best C points found in the embedding space. }

 $best\_emb \gets []$\\
 $Clusters \gets K\text{-}Means(\textrm{ $C$, $\overline{\mathcal{M}\text{atchSet}}$)}$\\
\For{cluster in Clusters}{ 

    $centroid_{norm} \gets\textrm{cluster centroid normalized to unit vector}$\\
    $pp \gets\textrm{ initialize first population with }centroid_{norm}$\\
    \For{generation in gen}{
    $pp'\gets \text{ generate new population based on } pp$\\
    $pp'_{norm} \gets\textrm{normalize to unit vector elements of } pp'$\\
    $loss \gets\textrm{[]}$\\
    \For{$pp'_{i}$ in $pp'_{norm}$ }{
        $loss \gets loss \cup 
        fitness(pp'_{i},cluster,\overline{\mathcal{D}\text{odgeSet}},th1,th2,\alpha,\beta,\gamma)$
    }
    $pp \gets $ top $|pp'|$ with lowest $loss$ from $pp'_{norm}$ 
} 
    $best\_emb \gets best\_emb \cup \textrm{embedding with lowest loss in } pp$
}
\Return $best\_emb$
\end{algorithm}


Our $DP loss$ is calculated for a target case $\overline{a}$ and a given set $\overline{S}$. It takes into account the number of cases in $\overline{S}$ that are farther away than the threshold to the case $\overline{a}$, and the sum of the distances between case $\overline{a}$ and all cases in $\overline{S}$. 

\begin{definition}[DP Loss]\label{def:DodgeCount_DistSum}
$loss$ is a normalized linear combination of $DodgeCount(\overline{a},\overline{S},th)$ and $DistSum(\overline{a},\overline{S})$: 
    \begin{align*}
            DP loss(\overline{a},\overline{S},th,m) = \\
             \frac{m \cdot \sum\limits_{\overline{s}\in \overline{S}} \mathds{1}_{\{Dist(\overline{a},\overline{s}) > th\}}(\overline{a}) + (1 - m) \cdot \sum\limits_{\overline{s} \in \overline{S} }^{} Dist(\overline{a},\overline{s})}{{\lvert \overline{S} \rvert}}
    \end{align*}

\noindent where $m\in [0,1]$ is the weight factor, and the indicator function $\mathds{1}_{\{Dist(\overline{a},\overline{s}) > th\}}(\overline{a})$ returns 1 when then condition $Dist(\overline{a},\overline{s}) > th$ is satisfied and 0 otherwise. 
\end{definition}

We normalize the result to ensure that the outcome is not related to the size of $\overline{S}$.
In our initial experiments, we observed that simply using the indicator function component was insufficient for the GA to select suitable individuals for the next generation. We solved that problem by incorporating the distances between the searched point and $\overline{S}$ members.



The DodgePersonation Fitness function of our GA is defined as a weighted sum of the positive $DP loss$ applied to each $\overline{\mathcal{M}\text{atchSet}_i}$ cluster and the negative $DP loss$ (i.e., multiplied by -1) applied to the $\overline{\mathcal{D}\text{odgeSet}}$ (cf. Definition~\ref{def:ga}). 



\begin{definition}[DodgePersonation Fitness Function] \label{def:ga}

Given a point $\overline{x}$ within the embedding space $\mathcal{E}$, let us consider $\overline{\mathcal{M}\text{atchSet}_i}$, a cluster of $\overline{\mathcal{M}\text{atchSet}}$, and the $\overline{\mathcal{D}\text{odgeSet}}$. Let $th1$ and $th2$ be the decision thresholds for $\overline{\mathcal{M}\text{atchSet}_i}$ and $\overline{\mathcal{D}\text{odgeSet}}$, respectively. Let $\alpha$ and $\beta$ be the weights used on the $DP loss$ components, and $\gamma$ be a weight parameter of the fitness function. The DodgePersonation Fitness function is defined as follows:
\begin{align*}
fitness(\overline{x},\overline{\mathcal{M}\text{atchSet}_i},\overline{\mathcal{D}\text{odgeSet}},th1,th2,\alpha,\beta,\gamma) = \\
\gamma * DPloss(\overline{x},\overline{\mathcal{M}\text{atchSet}_i},th1,\alpha) + \\
(1 - \gamma) * (-DPloss(\overline{x},\overline{\mathcal{D}\text{odgeSet}},th2,\beta))
\end{align*}

\end{definition}

We must highlight that our DodgePersonation Fitness function uses two different thresholds for the positive and negative $DPloss$ allowing them to be adjusted independently, which provides a more flexible method and more favorable results. 
Algorithm~\ref{alg:phase2} shows the One Face to Rule Them All Phase 1 pseudo-code for carrying out DodgePersonation Attack by searching a point or set of points in the embedding space. 

\subsection{Phase 2 - Attack Face Generation} 
Once we have obtained a set of points ($\overline{x_{i}}$) in the embedding space, we proceed to generate the corresponding Attack Face images. Starting with a Source Face $x$, our objective is to modify the image in such a way that, when passed through the FM, it is mapped closely to the Phase 1 point ($\overline{x_{i}}$). 



We developed a method for the FV task that builds upon any pre-selected source face image $x$, changing it as little as possible while simultaneously forcing the changed image to be matched to the previously obtained point $\overline{x_{i}}$. Drawing inspiration from renowned attack techniques like Projected Gradient Descent~\cite{madry2019deep}, the key idea of the Attack Face Generation method is to calculate the derivatives of the FM function with respect to the input image $x$ and alter the image to decrease the loss value, which is $Dist(FM(x),FM(\overline{x}_i))$. As far as we know, we are the first ones to apply this mapping of the embedding space into a specific image on the face image space, which gives full control over the initial identity in the face image.

\begin{algorithm}[ht]
\small
\SetAlgoLined
\KwInput{$x$: Source Face;\\
    \hspace{1cm}$\overline{y}$: point in the embedding space returned in Phase 1;\\
    \hspace{1cm}$\epsilon$: maximum change allowed for each pixel of the normalized image;\\
    \hspace{1cm}$iterations$: number of iterations;}
    
\KwOutput{$x_{adv}$: attack image}

\While{$x \neq MTCNN(x) $}{
$x \gets MTCNN(x)$\;
}

$x_{adv} \gets x$\;

\For{iter in $iterations$ }{ 
    $\overline{x}_{adv} \gets FM(x_{adv})$\;
    $loss \gets Dist(\overline{x}_{adv}, \overline{y}) $\;
    $gradients \gets \frac{\partial loss}{\partial x_{adv}}$\;
    $x_{adv} \gets Adam(x_{adv}, gradients)$\;
    $x_{adv} \gets clip(x_{adv}, [x_{adv}-\epsilon,x_{adv}+\epsilon])$\;
} 

\Return $x_{adv}$\;
\caption{One Face to Rule Them All Algorithm - Phase 2: Attack Face Generation.} \label{algo:derive}
\end{algorithm}

\begin{figure*}[htpb]
    \centering
    \includegraphics[width=0.99\textwidth]{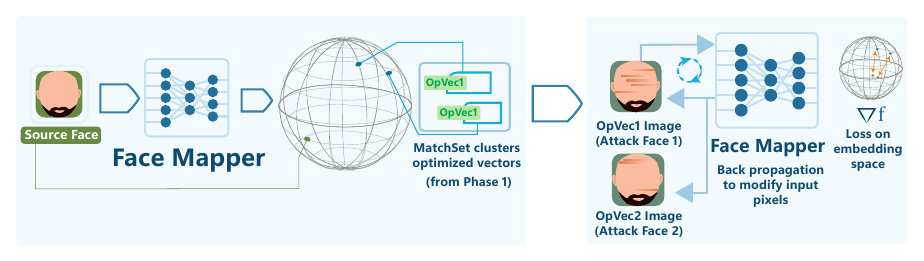}
    \caption{Overview of One Face to Rule Them All Phase 2: Attack Face Generation. The source image is mapped into the embedding space and modified to be in the proximity of a point determined in Phase 1. Using this method, we can conduct attacks on any given face image with control over the amount of change applied. }
    \label{fig:phase3}
\end{figure*}

Algorithm~\ref{algo:derive} shows the pseudo-code of the Attack Face Generation procedure. We first crop the face out from the Source Face $x$ using MTCNN~\cite{7553523} as the face detector. 
More precisely, we feed $x$ to MTCNN and resize it to match the input shape of the FM. This process is repeated until MTCNN outputs the exact same input without trimming, i.e., the input and output of MTCNN are of the same exact dimension. Then, the cropped version of $x$ is passed through a given number of iterations to embed the attack in the image. In each iteration, the algorithm evaluates the distance between the image mapped into the embedding space $FM(x)$ and the target point in the embedding space, $\overline{x}$. Then, the derivative of FM with respect to the image is calculated. Although the Adam optimizer~\cite{kingma2017adam} is typically used for updating neural network weights to minimize loss, we have repurposed the Adam optimizer to apply the obtained gradients to the image.
We also impose a constraint on the amount of change applied to control the difference ($\epsilon$) between the initial image and the altered one. This method provides a final modified version of $x$ that is visually very similar to the initial source image. An overview of the One Face to Rule Them All Phase 2 method is shown in Figure~\ref{fig:phase3}.


\section{Experimental Setup} \label{Evaluation}
In this section, we evaluate our methodology for deploying DodgePersonation Attacks by using the One Face to Rule Them All Algorithm. We first describe the dataset used in our experiments, followed by the implementation details of our experimental setting.

\subsection{Dataset}
We conducted our experiments on the LFW dataset~\cite{LFWTech}, which is one of the standard datasets for FV. This dataset contains 13,233 human images of 5,749 identities. We used the funneled version of the dataset. We obtained the FV decision threshold for a false acceptance rate\footnote{False Acceptance Rate, a metric commonly used to evaluate FV, represents the rate at which embeddings of different subjects are incorrectly matched as the same person.
} of 0.001 by using Euclidean distance on the LFW's provided training set that contains 1100 matching and 1100 mismatching pairs. Unless explicitly stated otherwise, the remaining experiments utilize a random subset of 5,749 images, with each identity in the dataset represented by a single image. The dataset, along with our code, is accessible from our GitHub repository.

\subsection{Evaluation Metrics}
The evaluation of the effectiveness of our approach involves calculating the coverage score on a given set of images $S$, given a set of Attack Faces $X$ as follows: $ coverage = \frac{|\{s \in S: \; \exists x \in X, \; Dist(FM(x), FM(s))<th\}|}{|S|}\time 100$. The coverage calculates the percentage of face images from $S$ that are matched, by at least one image in the set $X$ of Attack Faces. We calculate the coverage of the generated Attack Faces on both the $\mathcal{M}$atchSet and the $\mathcal{D}$odgeSet, which we aim to maximize and minimize, respectively.

\subsection{Implementation Details} \label{implement}
We randomly chose two disjoint sets of identities from the LFW dataset based on the given size of $\mathcal{M}$atchSet and $\mathcal{D}$odgeSet. 

One image per identity was chosen. 
If we use multiple images for each identity and assume that FM accurately groups these images in a nearby region of the embedding space, we are providing additional points to Phase 1 Embedding Space Search that are close to each other. This makes the task of the search algorithm to find an appropriate point in the embedding space easier. As a result, having this extra information allows the algorithm to produce more accurate outcomes. In contrast, when we opt to use only one image per identity, we are choosing a potentially more challenging situation.

We used the default settings for the MTCNN and resized the images into the input shape of the FM using the PIL library with BOX resampling algorithm. The input shape is 160x160, and the input images are normalized from $[0, 255]$ to $[-1, +1]$ as required by FaceNet~\cite{7298682}, the FM we used. FaceNet provides a mapping from the image space into a compact Euclidean Space and is trained on triplets of faces obtained from CASIA-WebFace dataset~\cite{yi2014learning} using triplet loss. Images are mapped onto the surface of a hyper-sphere, where each point in the embedding space has a fixed length of 512. 

For the LM-MA-ES~\cite{8410043} algorithm, we set the population size to 100 and the number of generations to 1000. If not specifically mentioned, the values for parameters $\alpha$, $\beta$, and $\gamma$ were set to 0.99, 0.99, and 0.9, respectively. Additionally, unless otherwise specified, the decision threshold values for th1 and th2 have been set to approximately 1.055, representing a false acceptance rate of 0.001. This particular false acceptance rate was chosen to replicate the testing conditions described in~\cite{shmelkin2021generating}. After the manipulation of each generation, all individuals were L2-normalized and assessed by the fitness function. The best 100 new individuals were chosen to replace the previous population.
    
For Phase 2, we used the Euclidean distance as the loss for mapping the output. The training iterations were set to 1,000. This value was obtained empirically after an initial set of experiments. 

\begin{figure*}[!h]
    \centering
    \includegraphics[width=0.85\textwidth]{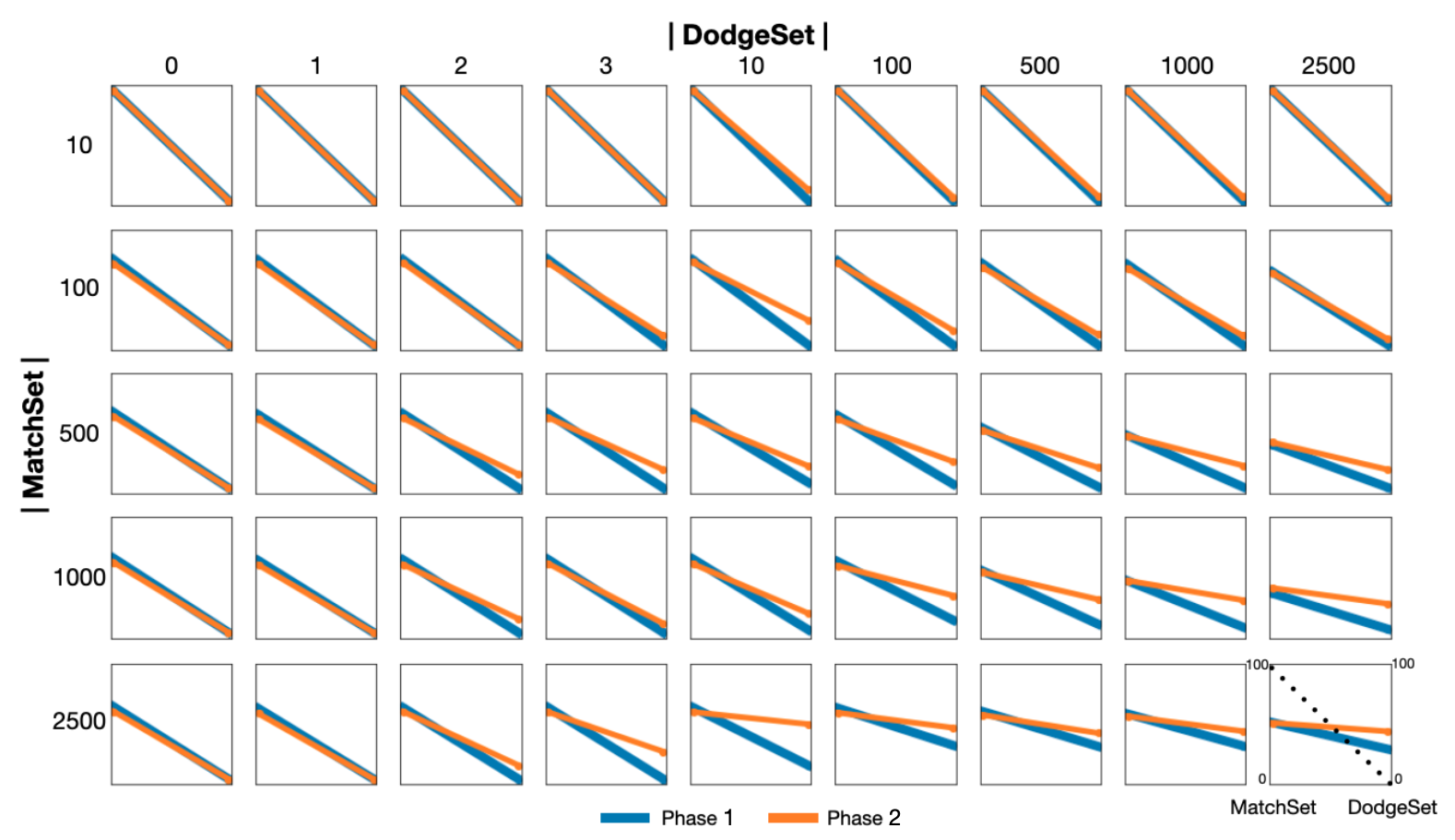}
    \caption{Coverage results for different $\mathcal{M}$atchSet and $\mathcal{D}$odgeSet sizes. Each plot shows the coverage on the $\mathcal{M}$atchSet on the left-hand side and the coverage on the $\mathcal{D}$odgeSet on the right-hand side. The coverage results on the two sets are connected by a line with a color corresponding to the phase where it was calculated.}
    \label{fig:phase3_10clusters}
\end{figure*}

\section{Results and Discussion} \label{Results}

This section presents the results of the experiments to assess the performance of our proposed One Face to Rule Them All Algorithm on the multiple scenarios defined in our taxonomy. 
We also present two ablation studies on parameters of our proposed DodgePersonation Fitness function. 


\subsection{Results for $\mathcal{M}$atchSet and $\mathcal{D}$odgeSet of varying size}\label{sec:res1}
In this set of experiments we considered 10 clusters in $\mathcal{M}$atchSet and evaluated our solution with $\mathcal{M}\text{atchSet}$ sizes of 10, 100, 500, 1000, 2500, and $\mathcal{D}\text{odgeSet}$ sizes of 0, 1, 2, 3, 10, 100, 1000, 2500. $\mathcal{M}$atchSet and $\mathcal{D}$odgeSet are created by randomly selecting elements from the LFW dataset. These sets are mapped into the embedding space to obtain $\overline{\mathcal{M}\text{atchSet}}$ and $\overline{\mathcal{D}\text{odgeSet}}$. We apply the One Face to Rule Them All Algorithm to find one point (Attack Face) in each cluster that satisfies the problem constraints.

We evaluated the coverage of the generated points in the embedding space at Phase 1 (in $\overline{\mathcal{M}\text{atchSet}}$ and $\overline{\mathcal{D}\text{odgeSet}}$) and the coverage of the corresponding generated Attack Faces at Phase 2 (in $\mathcal{M}$atchSet and $\mathcal{D}$odgeSet). This procedure is repeated 5 times. The average results are presented in Figure~\ref{fig:phase3_10clusters}. The coverage results of Phase 1 and Phase 2 are shown in blue and orange, respectively. Each subplot represents a specific scenario corresponding to a combination of a given $\mathcal{M}$atchSet and $\mathcal{D}$odgeSet size (e.g., the bottom right subplot shows the coverage results on a $\mathcal{M}$atchSet with size 2500 and a $\mathcal{D}$odgeSet of size 2500). In each subplot, the left side represents the coverage on $\overline{\mathcal{M}\text{atchSet}}$ (Phase 1 in blue) and $\mathcal{M}$atchSet (Phase 2 in orange) and the right side represents the coverage on the $\overline{\mathcal{D}\text{odgeSet}}$ (Phase 1 in blue) and $\mathcal{D}$odgeSet and (Phase 2 in orange). 
The optimal outcome for both Phase 1 and Phase 2 in each subplot is represented by a coverage of 100\% on the $\mathcal{M}$atchSet (left) and a coverage of 0\% on the $\mathcal{D}$odgeSet (right), which is a diagonal line drawn from the top left corner to the bottom right corner of the subplot (dashed line shown in the bottom right subplot).

As an example, let us focus on the subplot with a $\mathcal{M}$atchSet size of 100 and a $\mathcal{D}$odgeSet size of 2500. In this case, on Phase 1, 66.33\% of the $\overline{\mathcal{M}\text{atchSet}}$ is covered, while only 0.13\% of the $\overline{\mathcal{D}\text{odgeSet}}$ is covered. In Phase 2, we obtain a coverage of 65.33\% and 5.16\% on the $\mathcal{M}$atchSet and $\mathcal{D}$odgeSet, respectively. We observe that the results of Phase 1 and Phase 2 are very similar given the almost overlapping lines. This demonstrates the effectiveness of the One Face to Rule Them All Phase 2, which successfully generates Attack Faces of the selected identity that are close to the embedding space point found during Phase 1. 

Overall, we observe that the experiments with a lower number of faces to match and to dodge (subplots close to the top left) exhibit near-optimal results. As the number of faces to match and dodge increases, the coverage of the $\mathcal{M}$atchSet decreases, and the coverage of the $\mathcal{D}$odgeSet increases. The scenario with 2500 faces to match and 2500 faces to dodge (bottom right subplot) is the most challenging scenario and the least performing one. Still, even in this extreme scenario, a coverage of approximately 51\% is achieved on the $\mathcal{M}$atchSet on both phases, and a coverage of approximately 27\% and 43\% is achieved on the $\mathcal{D}$odgeSet on Phase 1 and 2, respectively. This is an excellent result given that we are able to match 1275 faces with only 10 attack images while dodging 1425 faces out of 2500. The detailed results are shown in Table~\ref{tab:phase3_10clusters} in Appendix~\ref{app:table}.

\begin{figure}[h]
    \centering
    \includegraphics[width=0.35\textwidth]{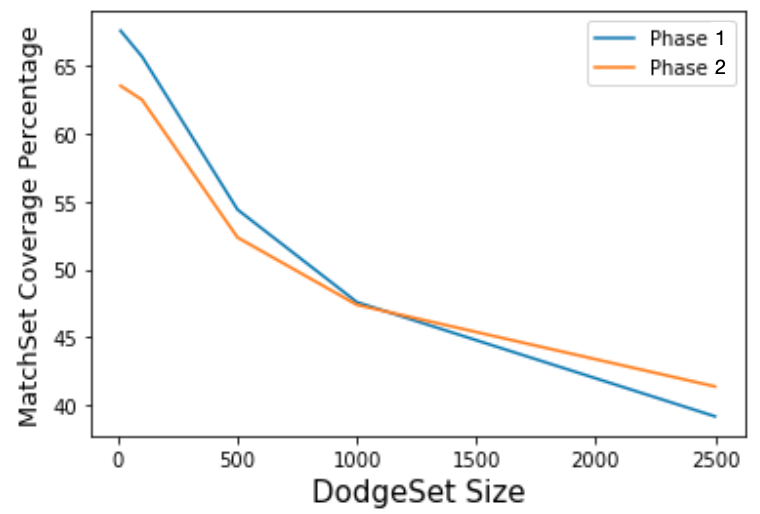}
    \caption{Coverage results of a fixed $\mathcal{M}$atchSet with 500 faces and different $\mathcal{D}$odgeSet sizes.}
    \label{fig:fixed_pset}
\end{figure}

We must highlight that Phase 2 tends to have lower performance when compared to Phase 1 results on the most challenging scenarios with higher sizes of $\mathcal{M}$atchSet and $\mathcal{D}$odgeSet, especially for the coverage of the $\mathcal{D}$odgeSet. 
Figure~\ref{fig:fixed_pset} illustrates this by examining the coverage results obtained with a fixed $\mathcal{M}$atchSet size of 500 and a $\mathcal{D}$odgeSet with size varying between 10 and 2500. The coverage of $\mathcal{M}$atchSet on both phases is higher when the size of $\mathcal{D}$odgeSet is lower, and it decreases with the increase of the $\mathcal{D}$odgeSet size. 


\subsection{Multi Identity Impersonation or Master Face} 

\subsubsection{Comparing One Face to Rule Them All algorithm against competitors}\label{sec:einstein}

\begin{figure*}[htpb]
    \centering
    \includegraphics[width=0.8\textwidth]{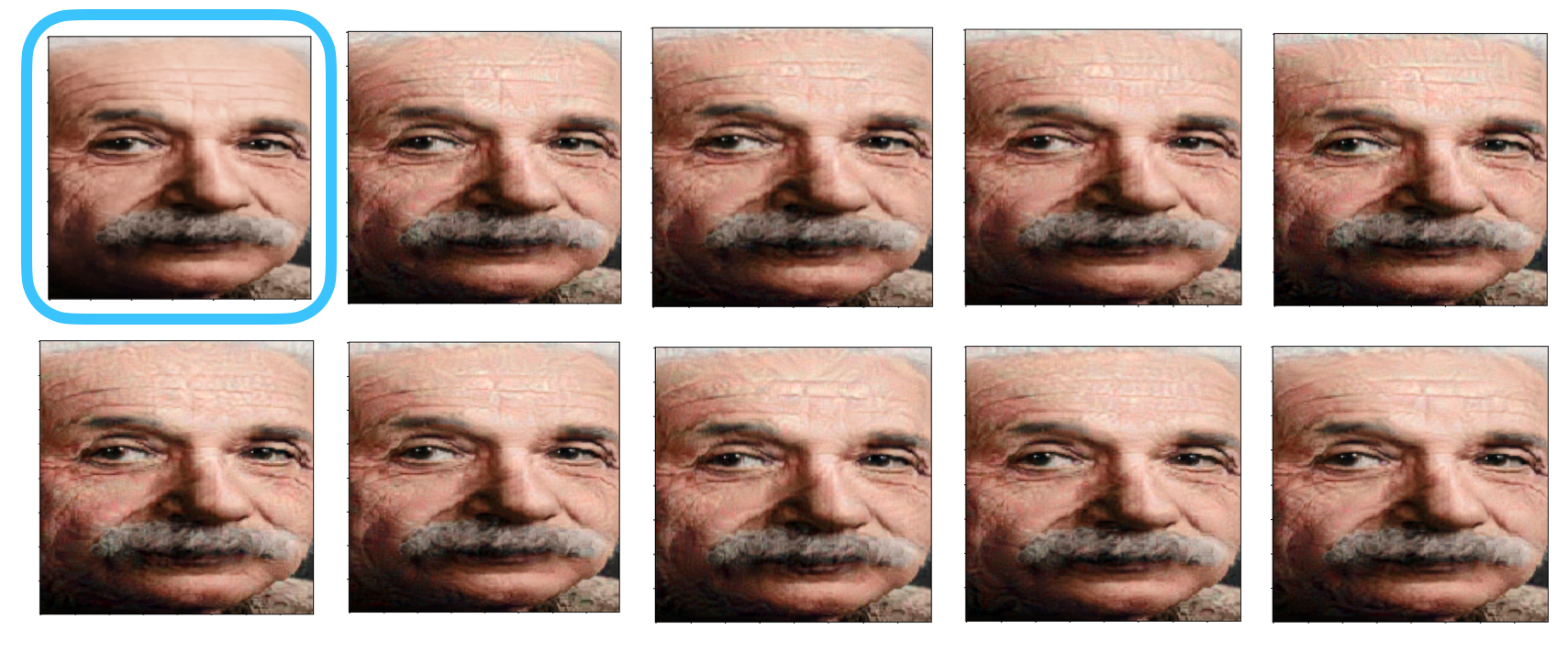}
    \caption{One Face to Rule Them All Algorithm for carrying out the Multi Identity Impersonation of 5749 identities using another image of Albert Einstein. The Original image (Source Face) is in a blue box. The remaining images are the Attack Faces, achieving a coverage of 58.5\% of the identities. The previous method covered only 43.82\% of the identities.
    }
    \label{fig:einstein}
\end{figure*}

An extreme scenario in Multi Identity Impersonation, also known as Master Face Attack, aims at impersonating a very large number of identities while considering an empty $\mathcal{D}$odgeSet. 
Recently, Shmelkin et al.~\cite{shmelkin2021generating} addressed the Master Face attack proposing a method to obtain a set of nine images generated using pre-trained GANs, which covered 43.82\% of the 5749 identities in the LFW dataset\footnote{In the context of the LFW dataset, the total of 5749 corresponds to the maximum number of unique identities that we can consider.}. However, the authors do not control the identity of the generated Master Faces. To compare our method with this attack, we recreated the precise testing setup in~\cite{shmelkin2021generating}, which involved using FaceNet trained on CASIA-WebFace with a decision threshold that corresponds to a false acceptance rate of 0.001, and a total of 5749 identities from the LFW dataset as the $\mathcal{M}$atchSet. We selected one image of Albert Einstein for this experiment, but any other Source Face could have been used. Figure~\ref{fig:einstein} shows the Attack Faces generated when applying our One Face to Rule Them All method. The original (Source Face) image is in a blue square, followed by the 9 Attack Faces generated. Our proposed One Face to Rule Them All method significantly improves the coverage achieved while allowing complete control over the identity of the nine generated images. We obtained a coverage of 58.5\% with the 9 images presented while the competing method achieved only a coverage of 43.82\% (also with 9 images but without any control over the identity of the generated images)\footnote{We present a second experiment carried out under the same conditions in Appendix~\ref{app:einstein} where we used a different Source Image of Albert Einstein and generated 9 Attack Faces with the One Face to Rule Them All method.}. This shows that our method can use any preferred Source Face, generating Attack Faces with state-of-the-art coverage. The Attack Faces generated have modifications that are almost imperceptible to humans, which strengthens the feasibility of the attack as it does not raise any alarms to humans. Moreover, the results show that the 9 images generated, although looking identical, cover different regions of the face space matching different identities.

\subsubsection{Analysis of the coverage of One Face to Rule Them All algorithm on Master Faces} \label{masterclustered}

\begin{figure}[htpb]
    \centering
    \includegraphics[width=0.45\textwidth]{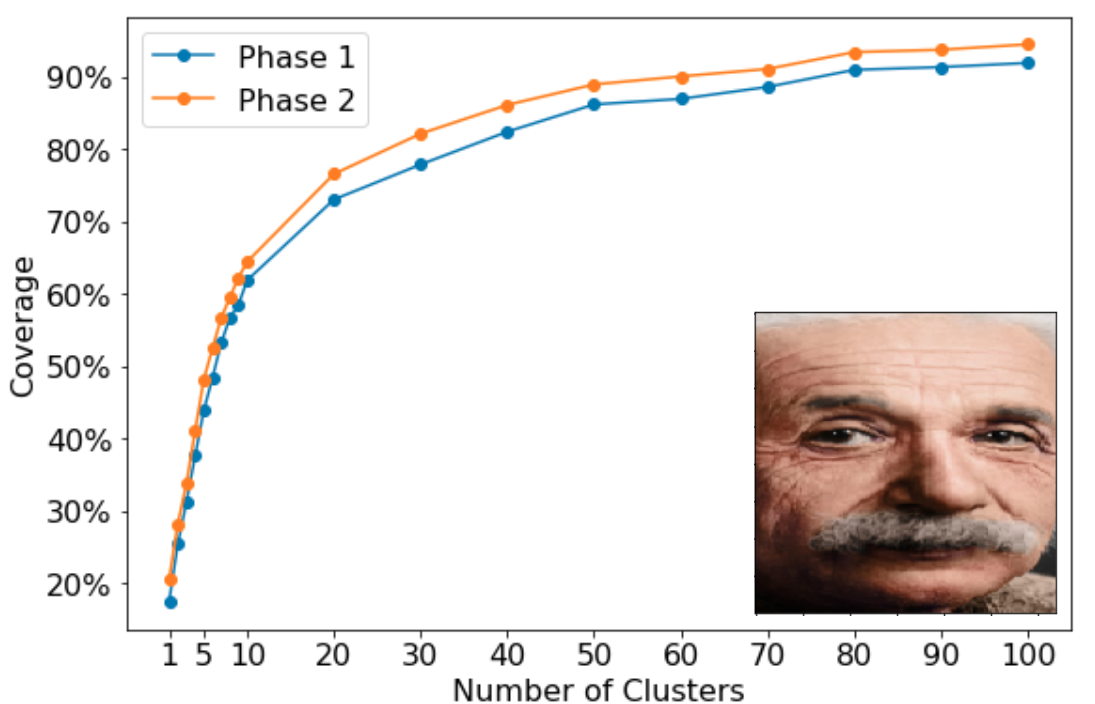}
    \caption{Coverage  results for different numbers of clusters (or Attack Faces) on a $\mathcal{M}$atchSet containing 5,749 identities from the LFW dataset. (Source Face used shown in the plot)}
    \label{fig:master}
\end{figure}

We carried out further experiments where we increased the number of Attack Faces generated to determine the number of images required to cover more than 90\% of the members in $\mathcal{M}$atchSet. The results of this experiment are presented in Figure~\ref{fig:master}. We observed that increasing the number of clusters, and consequently the number of Attack Faces, leads to higher coverage rates. Remarkably, with only 80 Attack Face images, the coverage exceeded 90\%. This experiment emphasizes the fact that with a relatively small number of images, we can efficiently cover a significant portion of the $\mathcal{M}$atchSet. This suggests that the FM might be clustering the images together in the embedding space. More importantly, our attack can successfully breach FV systems using a limited number of Attack Faces, highlighting the efficiency of our attack algorithm but also the fragility of these systems. The fact that with only 80 Attack Faces we match 90\% cases shows that FV systems are not safe, being breached with a small number of trials.

\subsubsection{Generalizability of One Face to Rule Them All Attack on Master Faces} 

The goal of this set of experiments is to assess the generalizability of the Attack Faces generated using the One Face to Rule Them All method for the Master Face Attack scenario. To assess this, we investigate the coverage of the Attack Faces on \textbf{other identities} that were not encountered during the generation process. We build a $\mathcal{M}$atchSet with a set of 2,500 identities randomly selected and craft a set of 10 Attack Faces to impersonate those identities. Then, we build a separate set of 2,500 identities, which we named $\mathcal{U}$nseenSet, containing only identities that were not present in the $\mathcal{M}$atchSet. Finally, the coverage of the 10 Attack Faces generated is evaluated on both $\mathcal{M}$atchSet and $\mathcal{U}$nseenSet. We repeated this test five times and report the average results. 

Our results show that with 10 Attack Faces, we achieve a coverage of 65.78\% and 60.22\% in the $\mathcal{M}$atchSet during Phase 1 and Phase 2, respectively. \textbf{Furthermore, the same 10 Attack Faces cover 58.25\% and 56.9\% of the $\mathcal{U}$nseenSet members in Phase 1 and Phase 2, respectively.} These findings provide strong evidence of the generalizability capability of our proposed One Face to Rule Them All method.


\subsection{Fitness Function ablation studies}

\subsubsection{Sensitivity of $\mathcal{D}$odgeSet threshold}\label{thresh}


Our proposed GA's DodgePersonation Fitness function (cf. Definition~\ref{def:ga}) takes into account two decision thresholds: $th1$ for $\mathcal{M}$atchSet and $th2$ for $\mathcal{D}$odgeSet. In our experiments,  $th1$ and $th2$ are both set to 1.055, as explained in Section~\ref{implement}. In this experiment, we investigate the impact of varying $th2$ on the coverage of $\mathcal{M}$atchSet and $\mathcal{D}$odgeSet.  

Our goal is two-fold: (i) understand if we can generate Attack Faces that are better at dodging the cases in $\mathcal{D}$odgeSet; and (ii) understand the impact of changing threshold $th2$ on the $\mathcal{M}$atchSet coverage. We hypothesize that if the GA is forced to consider a wider margin on the $\overline{\mathcal{D}\text{odgeSet}}$ by increasing $th2$, then more points can be dodged as the optimal points will be further away from this set members. 

We tested the initial value of $th2$ of 1.055 and included four other $th2$ values representing an increase of $th2$ by 3\%, 4\%, 5\%, and 6\%. We repeated these experiments 5 times using 1000 and 500 identities randomly selected for the $\mathcal{M}$atchSet and the $\mathcal{D}$odgeSet, respectively. The average coverage results on $\mathcal{M}$atchSet and $\mathcal{D}$odgeSet for Phase 1 and Phase 2 are displayed in Table~\ref{tab:nthresh}.


In Phase 1 and Phase 2, the coverage of the $\mathcal{D}$odgeSet when using the default value of 1.055 is 7.65\% and 30.20\%, respectively. Our special purpose GA was not able to avoid 7.65\% of the $\overline{\mathcal{D}\text{odgeSet}}$ members in the embedding space while the generated Attack Faces could not dodge 30.2\% of the $\mathcal{D}$odgeSet cases. 
Increasing the $\mathcal{D}$odgeSet $th2$ by 3\% led to a significant decrease in coverage percentages of the $\mathcal{D}$odgeSet in both phases, which confirms that increasing the threshold $th2$ helps to keep the Attack Faces further away from the $\mathcal{D}$odgeSet cases. However, we observe that the $\mathcal{M}$atchSet coverage is negatively impacted by this change, showing a decreased coverage. The $\mathcal{M}$atchSet coverage continues to decrease as the threshold $th2$ increases while the coverage of the $\mathcal{D}$odgeSet tends to zero. After an increase of 4\% in the $th2$ only the $\mathcal{M}$atchSet coverage is being affected because the $\mathcal{D}$odgeSet coverage is already very close to the ideal value of zero. Therefore, we confirm that adjusting the $\mathcal{D}$odgeSet threshold ratio can help to achieve better dodging results while experiencing lower impersonation results. This trade-off should be taken into account based on the problem's nature and the importance of dodging versus impersonation.

\begin{table}
\centering
\caption{Coverage results for different $\mathcal{D}$odgeSet thresholds $th2$ on Phases 1 and 2 (Phase 2 results in parenthesis). }\label{tab:nthresh}
\resizebox{0.4\textwidth}{!}{
\begin{tabular}{cccc}
\toprule
\multirow{2}{*}{$th2$}& \multirow{2}{*}{increase \%} & \multicolumn{2}{c}{Coverage }\\
\cmidrule{3-4}
&&
$\mathcal{M}$atchSet & $\mathcal{D}$odgeSet \\
\midrule
1.055&  -                     & 56.00 (54.68)                                         & 7.65 (30.20)                                      \\
1.086 & 3\%                & 41.56 (42.30)                                      & 0.00 (3.04)                                       \\
1.097 & 4\%                & 37.98 (39.50)                                      & 0.00 (1.48)                                       \\
1.107 & 5\%                 & 30.46 (35.62)                                        & 0.00 (0.60)                                       \\
1.118 & 6\%                 & 31.02 (33.72)                                      & 0.00 (0.40)                                       \\
\bottomrule
\end{tabular}
}

\end{table}

\subsubsection{Sensitivity of parameter $\gamma$} 
\begin{figure}[!t]
    \centering
    \includegraphics[width=0.3\textwidth]{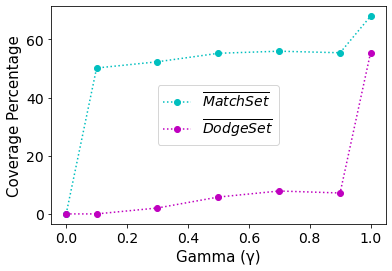}
    \caption{Impact of $\gamma$ on the coverage of $\overline{\mathcal{M}\text{atchSet}}$ and $\overline{\mathcal{D}\text{odgeSet}}$.}
    \label{fig:gamma}
\end{figure}

Parameter $\gamma$ in the GA Fitness function weights the $DPloss$ of the $\overline{\mathcal{M}\text{atchSet}}$ and $\overline{\mathcal{D}\text{odgeSet}}$. We randomly selected 1000 $\overline{\mathcal{M}\text{atchSet}}$ members and 500 $\overline{\mathcal{D}\text{odgeSet}}$ members and tested the values of 0, 0.1, 0.3, 0.5, 0.7, 0.9, and 1 for parameter $\gamma$. We repeated these experiments five times and reported the average. Figure~\ref{fig:gamma} shows the results of these experiments. 

We observe that, when $\gamma$ is 0, the focus is entirely on evading $\overline{\mathcal{D}\text{odgeSet}}$ members, leading to the neglect of $\overline{\mathcal{M}\text{atchSet}}$ members. On the other hand, when $\gamma$ is 1, the coverage of $\overline{\mathcal{M}\text{atchSet}}$ members peaks, while $\overline{\mathcal{D}\text{odgeSet}}$ members are ignored during dodging, resulting in a coverage of a significant percentage of $\overline{\mathcal{D}\text{odgeSet}}$ members. With an increase in $\gamma$ to 0.1, $\overline{\mathcal{M}\text{atchSet}}$ members gain considerable coverage while the coverage of $\overline{\mathcal{D}\text{odgeSet}}$ does not increase significantly. As $\gamma$ continues to grow, the emphasis on $\overline{\mathcal{M}\text{atchSet}}$ increases, resulting in coverage of more members, while the emphasis on $\overline{\mathcal{D}\text{odgeSet}}$ decreases, leading to less dodging of $\overline{\mathcal{D}\text{odgeSet}}$ members. Overall, the results are fairly stable for values of $\gamma$ between 0.1 and 0.9.  

\section{Conclusion}\label{sec:conc}
We proposed the DodgePersonation Attack for attacking FV systems. Our definition and taxonomy encompass both existing and novel types of attacks to FV systems. The DodgePersonation Attack aims at finding images that impersonate members in $\mathcal{M}$atchSet while avoiding the $\mathcal{D}$odgeSet members. We introduced a novel approach called One Face to Rule Them All, which successfully deploys the DodgePersonation Attack by generating adversarial images that impersonate a set of identities while dodging others. Our proposed algorithm achieves state-of-the-art performance in known types of attacks while also achieving outstanding results for novel attacks. Moreover, the One Face to Rule Them All algorithm generates attack images using a source face. This capability in our solution is not present in previous research. Finally, we must highlight that the generated attack images are built to embed the smallest change possible, and thus, the modifications applied can pass unnoticed to the human eye.


For future work, we plan to investigate the generalizability of our solution across different face descriptors. This could involve exploring the robustness of the solution to variations in the feature extraction process, which could have implications for the reliability of the system in real-world scenarios. The generalizability of our solution across different images of the same identity should also be explored. 
The exploration of our method on 3D images should also be considered, following a growing trend in related works~\cite{9963688}. A logical progression of this work involves experimenting with the printed version of the Attack Faces to carry out Physical or Presentation Attacks. 
Finally, we will consider the extension of the proposed algorithm to fingerprint recognition systems and investigate its effectiveness in generating a master fingerprint set.

\bibliographystyle{plain}
\bibliography{refs}

\appendix

\section{Appendices}

\subsection{Single Impersonation and Single Dodging Scenario}

For this study, we randomly pick a single picture from $\mathcal{F}$ as the $\mathcal{M}$atchSet (with an empty $\mathcal{D}$odgeSet) and aim to modify Albert Einstein's face image to resemble the identity in the selected picture. The experiment was conducted ten times, and the outcome revealed 100\% coverage, implying that in all ten trials, the modified image of Einstein impersonated the randomly chosen person successfully.

In order to address a single dodging scenario, we randomly select an image from $\mathcal{F}$ as the $\mathcal{D}$odgeSet (with an empty $\mathcal{M}$atchSet) and alter it in order to dodge or avoid its identity. Therefore, the image used for the attack is the only member of $\mathcal{D}$odgeSet. The experiment is repeated ten times, and it is found that in ten out of the ten attempts, the dodging was successful.

\subsection{Discussion about the $\mathcal{M}$atchSet and $\mathcal{D}$odgeSet on the DodgePersonation Attack}

The definition we propose requires a $\mathcal{M}$atchSet and a $\mathcal{D}$odgeSet. One may argue that the use of $\mathcal{D}$odgeSet might be redundant by observing that when matching the identity or identities in $\mathcal{M}$atchSet any other images will be dodged automatically. However, we claim that this might not be true, and thus the use of both $\mathcal{M}$atchSet and $\mathcal{D}$odgeSet is necessary. Let us illustrate this with an example. 
Suppose we face the challenge of creating an image that appears as person A to humans, is recognized by the FV system as person B, and avoids identification as person C (or dodges person C). While some argue that if the FV system confirms the image as person B, it automatically avoids other identities like person C, this assumption is not foolproof. FV systems are imperfect, as demonstrated by the presence of "master faces", which means that a solution may match more identities than those in the MatchSet. To address this issue, a DodgeSet is necessary to ensure the robustness and reliability of the system, explicitly considering unwanted identities.

\subsection{Exploring the Distribution of Identities in the Embedding Space}
\begin{figure}[htpb]
    \centering
    \includegraphics[width=0.45\textwidth]{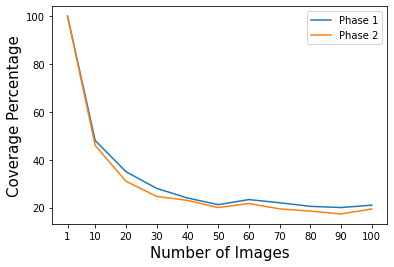}
    \caption{Coverage percentage by the number of identities to impersonate using a single image.}
    \label{fig:impersonation_variation}
\end{figure}

Two experiments were performed to gain a better understanding of how the face images of different identities are distributed in the embedding space. The first study involves selecting a random image as the $\mathcal{M}$atchSet (with an empty $\mathcal{D}$odgeSet) and then modifying Albert Einstein's photo to impersonate it. This is repeated five times, then the average result is reported. The process is then repeated for two images, then three, and so on, up to ten images, each time repeating the experiment five times. The outcomes are displayed in Figure~\ref{fig:impersonation_variation}. It is evident that as the number of identities to impersonate increases, the coverage percentage decreases. This suggests that the embeddings are not localized within a specific region of the embedding space, making it impossible for a single point to accurately cover all of them.

\subsection{Detailed results of varying size of $\mathcal{M}$atchSet and $\mathcal{D}$odgeSet}\label{app:table}

Table~\ref{tab:phase3_10clusters} shows the detailed results of the experiment carried out in Section~\ref{sec:res1} and summarized in Figure~\ref{fig:phase3_10clusters}.

\begin{table}
\centering
\caption{Results of different $\mathcal{M}$atchSet and $\mathcal{D}$odgeSet sizes in Phase 1 and Phase 2 of the One Face to Rule Them All Algorithm. The results are the average of 5 runs for 10 clusters. This table corresponds to the results shown in Figure~\ref{fig:phase3_10clusters}.}
\label{tab:phase3_10clusters}

\resizebox{\columnwidth}{!}
{

\begin{tabular}{@{}rrrrrr@{}}
\toprule
 \multirow{2}{*}{$|\mathcal{M}$atchSet$|$} &   \multirow{2}{*}{$|\mathcal{D}$odgeSet$|$} &  \multicolumn{4}{c}{Coverage}\\
  \cline{3-6}
\vspace*{-0.25cm}\\  
  & &  $\overline{\mathcal{M}\text{atchSet}}$ &  $\overline{\mathcal{D}\text{odgeSet}}$ &  $\mathcal{M}$atchSet &  $\mathcal{D}$odgeSet \\
\midrule
         10 &           0 &    100.00 &      0.00 &            100.00 &              0.00 \\ \hline
         10 &           1 &    100.00 &      0.00 &            100.00 &              0.00 \\ \hline
         10 &           2 &    100.00 &      0.00 &            100.00 &              0.00 \\ \hline
         10 &           3 &    100.00 &      0.00 &            100.00 &              0.00 \\ \hline
         10 &          10 &    100.00 &      0.00 &            100.00 &             10.00 \\ \hline
         10 &         100 &    100.00 &      0.00 &            100.00 &              3.25 \\ \hline
         10 &         500 &    100.00 &      0.00 &            100.00 &              4.15 \\ \hline
         10 &        1000 &    100.00 &      0.00 &            100.00 &              3.48 \\ \hline
         10 &        2500 &    100.00 &      0.45 &            100.00 &              2.48 \\ \hline
        100 &           0 &     77.75 &      0.00 &             74.00 &              0.00 \\ \hline
        100 &           1 &     77.00 &      0.00 &             73.50 &              0.00 \\ \hline
        100 &           2 &     77.50 &      0.00 &             74.50 &              0.00 \\ \hline
        100 &           3 &     77.75 &      0.00 &             74.50 &              8.33 \\ \hline
        100 &          10 &     77.25 &      0.00 &             75.25 &             22.50 \\ \hline
        100 &         100 &     76.25 &      0.00 &             74.25 &             13.25 \\ \hline
        100 &         500 &     73.75 &      0.05 &             70.25 &              9.35 \\ \hline
        100 &        1000 &     73.00 &      0.05 &             69.50 &              8.42 \\ \hline
        100 &        2500 &     66.33 &      0.13 &             65.33 &              5.16 \\ \hline
        500 &           0 &     68.85 &      0.00 &             65.15 &              0.00 \\ \hline
        500 &           1 &     66.50 &      0.00 &             62.55 &              0.00 \\ \hline
        500 &           2 &     67.40 &      0.00 &             63.35 &             12.50 \\ \hline
        500 &           3 &     67.70 &      0.00 &             64.10 &             16.67 \\ \hline
        500 &          10 &     67.55 &      5.00 &             63.50 &             20.00 \\ \hline
        500 &         100 &     65.65 &      3.50 &             62.47 &             24.00 \\ \hline
        500 &         500 &     54.40 &      1.15 &             52.35 &             18.55 \\ \hline
        500 &        1000 &     47.60 &      0.88 &             47.40 &             20.38 \\ \hline
        500 &        2500 &     39.20 &      0.45 &             41.40 &             16.67 \\ \hline
       1000 &           0 &     67.62 &      0.00 &             63.82 &              0.00 \\ \hline
       1000 &           1 &     65.42 &      0.00 &             61.52 &              0.00 \\ \hline
       1000 &           2 &     66.12 &      0.00 &             61.78 &             12.50 \\ \hline
       1000 &           3 &     66.45 &      0.00 &             62.68 &              8.33 \\ \hline
       1000 &          10 &     66.85 &      2.50 &             62.38 &             17.50 \\ \hline
       1000 &         100 &     64.47 &     11.33 &             60.70 &             33.67 \\ \hline
       1000 &         500 &     56.00 &      7.65 &             54.68 &             30.20 \\ \hline
       1000 &        1000 &     47.00 &      5.00 &             47.30 &             29.50 \\ \hline
       1000 &        2500 &     35.63 &      3.15 &             40.53 &             26.15 \\ \hline
       2500 &           0 &     65.86 &      0.00 &             61.56 &              0.00 \\ \hline
       2500 &           1 &     64.23 &      0.00 &             59.57 &              0.00 \\ \hline
       2500 &           2 &     64.88 &      0.00 &             61.31 &             12.50 \\ \hline
       2500 &           3 &     65.11 &      0.00 &             60.92 &             25.00 \\ \hline
       2500 &          10 &     65.16 &     12.50 &             61.01 &             50.00 \\ \hline
       2500 &         100 &     64.61 &     30.67 &             60.36 &             46.67 \\ \hline
       2500 &         500 &     61.47 &     29.75 &             58.34 &             41.85 \\ \hline
       2500 &        1000 &     59.25 &     30.20 &             57.29 &             43.60 \\ \hline
       2500 &        2500 &     51.42 &     27.24 &             51.20 &             43.40 \\ 
\bottomrule
\end{tabular}

}
\end{table}

\subsection{Extra Results on Multi Identity Impersonation or Master Face}\label{app:einstein}

\begin{figure*}[b]
    \centering
    \includegraphics[width=0.8\textwidth]{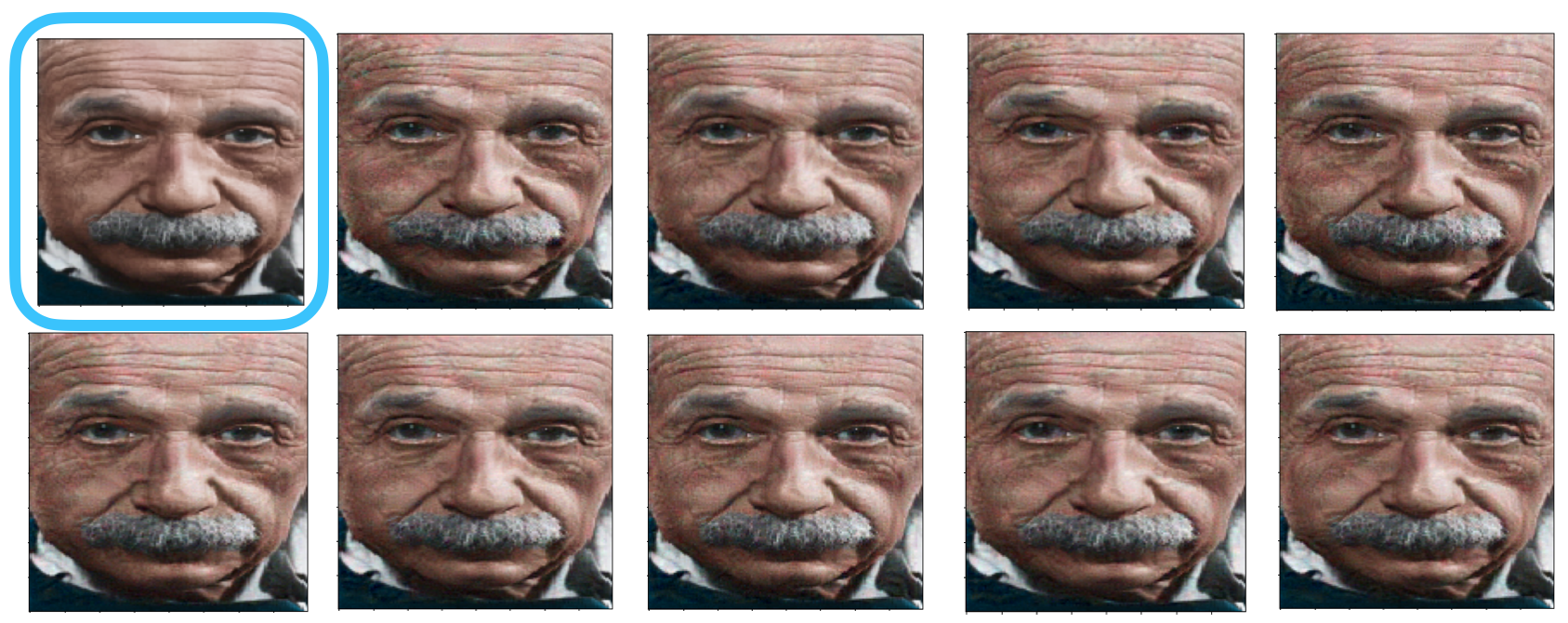}
    \caption{One Face to Rule Them All Algorithm for carrying out the Multi Identity Impersonation of 5749 identities using the image of Albert Einstein. The Original image (Source Face) is in a blue box. The remaining images are the Attack Faces, achieving a coverage of 57.27\% of the identities. The previous method covered only 43.82\% of the identities.
    }
    \label{fig:einstein2}
\end{figure*}

Figure~\ref{fig:einstein2} shows the results of an experimental setting similar to the one described in Section~\ref{sec:einstein} but using a different Source Face image. In this case, we considered a second image of Albert Einstein, which achieved 57.27\% of coverage.

\end{document}